\documentclass[12pt,journal,draftclsnofoot,a4paper,oneside,onecolumn]{IEEEtran}
\usepackage{times,amsmath,epsfig}
\usepackage{cite}
\usepackage{graphicx}
\usepackage{psfrag}
\usepackage{subfigure}
\usepackage{url}
\usepackage{stfloats}
\usepackage{amsmath}
\usepackage{amssymb}
\usepackage{array}

\newcounter{step}
\newlength{\totlinewidth}
  {\end{list}%
  \rule{\linewidth}{1pt}}
\newcounter{substep}

  {\end{list}}

\newlength{\aligntop}
\newlength{\alignbot}
 \makeatletter
\renewenvironment{align}{%
  \vspace{\aligntop}
  \start@align\@ne\st@rredfalse\m@ne
}{%
  \math@cr \black@\totwidth@
  \egroup
  \ifingather@
    \restorealignstate@
    \egroup
    \nonumber
    \ifnum0=`{\fi\iffalse}\fi
  \else
    $$%
  \fi
  \ignorespacesafterend%
  \vspace{\alignbot}\par\noindent
} \makeatother

\title{Joint Relay Selection and Analog Network Coding using Differential Modulation in Two-Way Relay Channels}

\author{Lingyang~Song,~\IEEEmembership{Member,~IEEE,} Guo Hong,~\IEEEmembership{Member,~IEEE,} Bingli~Jiao,~\IEEEmembership{Member,~IEEE,} and Merouane
Debbah,~\IEEEmembership{Senior Member,~IEEE}
\thanks{Copyright (c) 2010 IEEE. Personal use of this material is permitted. However, permission to use this material for any other purposes must be obtained from the IEEE by sending a request to pubs-permissions@ieee.org.}
\thanks{Lingyang~Song and Bingli~Jiao are with Peking University, China (e-mail:
\protect\url{lingyang.song@pku.edu.cn,jiaobl@pku.edu.cn}).}
\thanks{Guo Hong is with Institute of Applied Physics and Computational Mathematics, China (e-mail: \protect\url{guo
hong@iapcm.ac.cn}).}
\thanks{Merouane Debbah is SUP¨¦LEC, Alcatel-Lucent Chair in Flexible Radio, 3 rue Joliot-Curie, FR-91192 Gif Sur Yvette, France (e-mail: \protect\url{merouane.debbah@supelec.fr}).}
}

\begin{document}

\maketitle
\begin{abstract}
In this paper, we consider a general bi-directional relay network
with two sources and $N$ relays when neither the source nodes nor
the relays know the channel state information (CSI). A joint relay
selection and analog network coding using differential
modulation~(RS-ANC-DM) is proposed. In the proposed scheme, the two
sources employ differential modulations and transmit the
differential modulated symbols to all relays at the same time. The
signals received at the relay is a superposition of two transmitted
symbols, which we call the analog network coded symbols. Then a
single relay which has minimum sum SER is selected out of $N$ relays
to forward the ANC signals to both sources. To facilitate the
selection process, in this paper we also propose a simple sub-optimal
Min-Max criterion for relay selection, where a single relay which
minimizes the maximum SER of two source nodes is selected. Simulation results show that the proposed Min-Max
selection has almost the same performance as the optimal selection,
but is much simpler. The performance of the proposed RS-ANC-DM
scheme is analyzed, and a simple asymptotic SER expression is
derived. The analytical results are verified through simulations.
\end{abstract}

\pagebreak

\section{Introduction}
In a bi-directional relay network, two source nodes exchange their
messages through the aid of one or multiple relays. The transmission
in bi-directional relay network can take place over either four,
three or two time slots. In the four time slots transmission
strategy, the relay helps to forward source $S_1$'s message to
source $S_2$ in the first two time slots and source $S_2$'s message
to source $S_1$ in the next two time slots. Four time slots
transmission has been shown to be very inefficient. When the relay
receives two sources' messages, it combines them before forwarding
to the destination, which will save one time slot transmission. This
three time slots transmission scheme is usually referred to as the
digital network coding
\cite{Ahlswede2000}--\hspace{-.00001mm}\cite{Liu2008}. In this
method, two source nodes transmit to the relay, separately. The
relay decodes the received signals, performs binary network coding,
and then broadcasts it back to both source nodes.

To further improve the spectral efficiency, the message exchange
between two source nodes can actually take place in two time slots.
In the first time slot, both source nodes transmit at the same time
so that the relay receives a superimposed signal. The relay then
amplifies the received signal and broadcasts it to both source nodes
in the second time slots. This scheme is referred to as the analog
network coding (ANC)
\cite{Popovski2007}--\hspace{-.00001mm}\cite{Katti2007}. Various
transmission schemes and wireless network coding schemes in
bi-directional relay networks have been analyzed and compared in
\cite{Eslamifar2010}--\hspace{-.00001mm}\cite{Oechtering2008}.

Most of existing works in bi-directional relay communications
consider the coherent detection at the destination and assume that
perfect channel state information~(CSI) are available at the sources
and relays
\cite{Ahlswede2000}--\hspace{-.00001mm}\cite{Oechtering2008}. In
some scenarios, e.g. the fast fading environment, the acquisition of
accurate CSI may become difficult. In this case, the non-coherent or
differential modulation would be a practical solution. In a
differential bi-directional relay network, each source receives a
superposition of differentially encoded signals from the other
source, and it has no knowledge of CSI of both channels. All these
problems present a great challenge for designing differential
modulation schemes in two-way relay channels.

To solve this problem, in \cite{Tao2008}, a non-coherent receiver
for two-way relaying was proposed for ANC based bi-directional relay
networks. However, the schemes result in more than 3 dB performance
loss compared to the coherent detection. To further improve the
system performance, a differential ANC scheme was proposed in
\cite{Song} and a simple linear detector was developed to recover
the transmitted signals at two source nodes. Results have shown that
it only has about 3~dB performance loss compared to its coherent
counterpart.

Recently, it has been shown that the performance of wireless relay
networks can be further enhanced by properly selecting the relays
for transmission \cite{Zhao2006}--\hspace{-.00001mm}\cite{Song2009}.
Consequently, it is beneficial to design an effective relay
selection scheme for the bi-directional transmission scheme with
multiple relays as well in order to achieve spatial diversity. In
this paper, we propose a bi-directional joint relay selection and
analog network coding using differential modulation~(RS-ANC-DM) so
that the CSI is not required at both sources and the selected relay.
In the proposed BRS-DS-ANC scheme, two source nodes first
differentially encoder their messages and then broadcast them to all
the relays at the same time. The signals received at the relay is a
superposition of two transmitted symbols. Then a single relay which
minimizes the sum SER of two source nodes is selected out of all
relays to forward the ANC signals to both sources. Each source node
then performs the differential detection and subtract its own
message to recover the message transmitted by the other source node.

The performance of optimal relay selection is very difficult to
analyze. To facilitate the analysis and selection procedure, in this
paper we also propose a simple sub-optimal Min-Max criterion for
relay selection, where a single relay which minimizes the maximum
BER of two source nodes is selected. Simulation results show that
the proposed Min-Max selection has almost the same performance as
the optimal selection, but is much simpler. The performance of the
proposed BRS-ANC scheme is analyzed, and an asymptotic SER
expression is derived. The analytical results are verified through
simulations.

The rest of the paper is organized as follows: In Section~II, we
describe the proposed BRS-DS-ANC scheme. The performance of the
RS-ANC-DM is analyzed Section~III. Simulation results are provided
in Section~IV. In Section~V, we draw the conclusions.

\emph{\textbf{Notation}}: Boldface lower-case letters denote
vectors, $(\cdot)^{*}$, $(\cdot)^{T}$ and $(\cdot)^{H}$ represent
conjugate, transpose, and conjugate transpose, respectively.
$\mathbb{E}$ is used for expectation, $\texttt{Var}$ represents
variance, $\|\textbf{x}\|^2=\textbf{x}^H\textbf{x}$, and
$\mathfrak{R}(\cdot)$ denotes real part.

\section{Joint Relay Selection and Analog Network Coding using Differential Modulation}
\begin{figure}[]
\centering
\includegraphics[height=7.5in,width=5.5in]{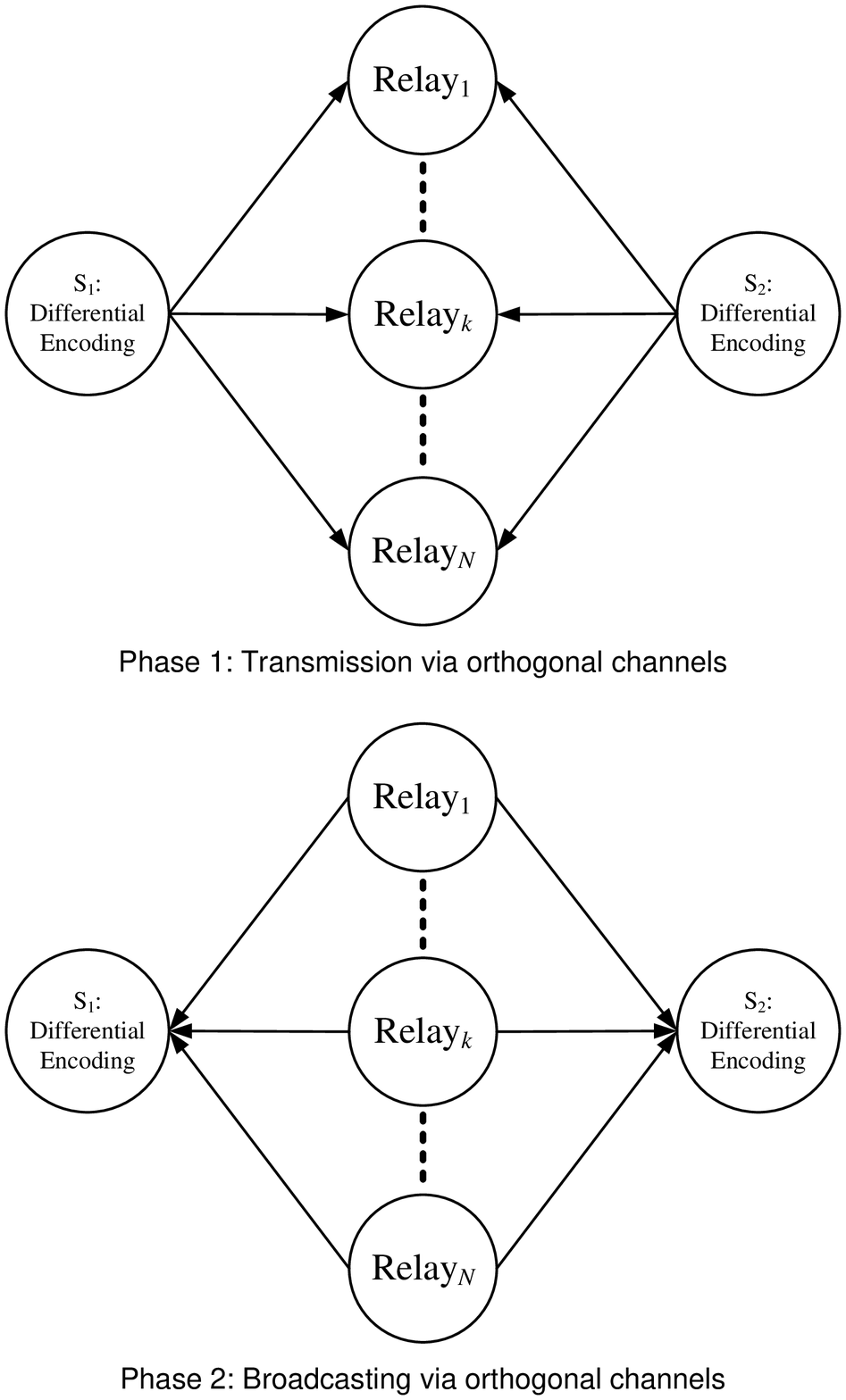}
\caption{Block diagram of the proposed BRS-DANC scheme.}
\label{Fig:Diff_bidirectional_relay_dia}
\end{figure}

We consider a general bi-directional relay network, consisting of
two source nodes, denoted by $S_1$ and $S_2$, and $N$ relay nodes,
denoted by $R_1,\ldots,R_N$. We assume that all nodes are equipped
with single antenna. In the proposed RS-ANC-DM scheme, each message
exchange between two source nodes takes place in two phases, as
shown in Fig.~\ref{Fig:Diff_bidirectional_relay_dia}. In the first
phase, both source nodes simultaneously send the differentially
encoded information to all relays and the signal received at each
relay is a superimposed signal. In the second phase, an optimal
relay node is selected to forward the received signals to two source
nodes and all other relay nodes keep idle.


\subsection{Differential Encoding and Decoding in Two-Way Relay Channels}

Let $c_i(t)$, $i=1, 2$, denote the symbol to be transmitted by the
source $S_i$ at the time $t$. We consider a MPSK modulation and
assume that $c_i(t)$ is chosen from a MPSK constellation of unity
power $\mathcal{A}$. Source $i$ first differentially encodes the
information symbols $c_i(t)$
\begin{equation}
    s_i(t)=s_i(t-1){c_i(t)}
 \label{Eq:s1}
\end{equation}

The differential encoded signals are then simultaneously transmitted
by two source nodes with unit transmission power to all the
relays. The signal received in the $k$-th relay at time $t$ can be
expressed as
\begin{equation}
    y_{r,k}(t)=h_{1,k}s_1(t)+h_{2,k}s_2(t)+n_{r,k}(t),
 \label{Eq:yr}
\end{equation}
where $h_{i,k}$, $i=1,2,k=1,...,N$, is the fading coefficient
between $S_i$ and $R_k$. In this paper, we consider a quasi-static
fading channel for which the channels are constant within one frame,
but change independently from one frame to another. $n_{r,k}(t)$ is
a zero mean complex Gaussian random variable with two sided power
spectral density of $N_0/2$ per dimension.


Upon receiving the signals, the relay $R_k$ then processes the
received signal and then forwards to two source nodes. Let
$x_{r,k}(t)$ be the signal generated by the relay $R_k$ and it is
given by
\begin{equation}
    x_{r,k}(t)=\beta_ky_{r,k}^*(t),
 \label{Eq:xr}
\end{equation}
where $\beta_k$ is an amplification factor, so that the signal
transmitted by the relay satisfy the following power constraint
\begin{equation}
    \mathbb{E}(|x_{r,k}(t)|^2)=1.
 \label{Eq:pr}
\end{equation}

We should note that unlike the traditional ANC schemes
\cite{Popovski2007}--\hspace{-.00001mm}\cite{Katti2007}, the relay
forwards the conjugate of the received signal. The reason of doing
this is to facilitate the differential detection at the destination
\cite{Song}.

Substituting Eqs. (\ref{Eq:yr}) and (\ref{Eq:xr}) into
(\ref{Eq:pr}), we can derive $\beta_k$,
\begin{equation}
    \beta_k=\sqrt{\frac{1}{|h_{1,k}|^2+|h_{2,k}|^2+N_0}}
 \label{Eq:beta}
\end{equation}

However, as the relay has no CSI, $\beta_k$ has to be obtained in
other way. Let
$\textbf{y}_{r,k}=[{y}_{r,k}(1),\ldots,{y}_{r,k}(L)]^T$,
$\textbf{s}_1=[{s}_1(1),\ldots,{s}_1(L)]^T$,
$\textbf{s}_2=[{s}_2(1),\ldots,{s}_2(L)]^T$,
$\textbf{n}_{r,k}=[{n}_{r,k}(1),\ldots,{n}_{r,k}(L)]^T$, where $L$
is the frame length. Then we can rewrite the received signals in
(\ref{Eq:yr}) in a vector format as follows
\begin{align}
    \textbf{y}_{r,k}=h_{1,k}\textbf{s}_1+h_{2,k}\textbf{s}_2+\textbf{n}_{r,k},
 \label{Eq:yrvec}
\end{align}
and $\beta_k$ can be then approximated by the $k$-th relay node as
\begin{align}
    \beta_k=\sqrt{\frac{\mathbb{E}\{\textbf{y}_{r,k}^H\textbf{y}_{r,k}\}}{L}}\approx\sqrt{\frac{{\textbf{y}_{r,k}^H\textbf{y}_{r,k}}}{L}},
 \label{Eq:beta}
\end{align}

%

After deriving $\beta_k$, the relay $R_k$ then forwards $x_{r,k}(t)$
to two source nodes. Since $S_1$ and $S_2$ are mathematically
symmetrical, for simplicity, in the next we only discuss the
decoding as well as the analysis for signals received by $S_1$. The
signal received by $S_1$ at time $t$, denoted by $y_{i,k}(t)$, can
be written as
\begin{align}
    y_{1,k}(t)&={\beta_k}h_{1,k}y_{r,k}^*(t)+n_{i,k}(t)
    \nonumber \\
          &={\mu_k}s_1^*(t)+{\nu_k}s_2^*(t)+w_{i,k}(t) \nonumber \\
          &={\mu_k}s_1^*(t)+{\nu_k}s_2^*(t-1)c_2^*(t)+w_{1,k}(t),
 \label{Eq:y1}
\end{align}
where $\mu_k\triangleq{\beta_k}|h_{1,k}|^2>{0}$,
$\nu_k\triangleq{\beta_k}h_{1,k}h_{2,k}^*$, and
$w_{1,k}(t)\triangleq{\beta_k}h_{1,k}n_{r,k}^*(t)+n_{1,k}(t)$.

%


Since $s_1(t)$ is known to the source $S_1$, to decode $c_2(t)$,
$S_1$ needs to estimate $\mu_k$ and $\nu_k$. Let
$\textbf{y}_{1,k}=[{y}_{1,k}(1),\ldots,{y}_{1,k}(L)]^T$ and
$\textbf{w}_{1,k}=[{w}_{1,k}(1),\ldots,{w}_{1,k}(L)]^T$. Then at
high SNR, we can obtain the following approximation
\begin{equation}
    \mu_k^2+|\nu|_k^2\approx{\frac{\textbf{y}_{1,k}^H\textbf{y}_{1,k}}{L}}.
 \label{Eq:uv}
\end{equation}

Since the source node $S_1$ can retrieve its own information
$s_1(t-1)$ and $c_1(t)$, we have
\begin{align}
    \widetilde{y}_{1,k}(t)&\triangleq{c_1^*(t)}y_{1,k}(t-1)-y_{1,k}(t)
        \nonumber \\
          &={\nu_k}s_2^*(t-1)\left(c_1(t)-c_2(t)\right)^*
    +\widetilde{w}_{1,k}(t),
 \label{Eq:y1y0}
\end{align}
where
$\widetilde{w}_{1,k}(t)\triangleq{{c_1(t)}w_{1,k}(t-1)+w_{1,k}(t)}$.
Then, $|\nu_k|^2$ can be approximated as \cite{Song}
\begin{align}
    |\nu_k|^2\approx\frac{\widetilde{\textbf{y}}_{1,k}^H\widetilde{\textbf{y}}_{1,k}}
    {L\mathbb{E}\left[|c_1(t)-c_2(t)|^2\right]},
 \label{Eq:nu}
\end{align}
where
$\widetilde{\textbf{y}}_{1,k}=[\widetilde{y}_{1,k}(1),\ldots,\widetilde{y}_{1,k}(L-1)]^T$,
and $\mathbb{E}[|c_1(t)-c_2(t)|^2]$ is a constant which can be
pre-calculated by two source nodes, which is given in Appendix~A. As $\mu_k$ is positive, it can
be derived from (\ref{Eq:uv}) and (\ref{Eq:nu}) as
\begin{align}
    \mu_k\approx{\left(\Theta_k\right)_{+}} \label{Eq:muf}
\end{align}
where $\Theta_k\triangleq
{\frac{{\textbf{y}_{1,k}^H\textbf{y}_{1,k}}}{L}-|\nu_k|^2}$,
$(X)_{+}$ is equal to $X$ when $X\geq{0}$ and otherwise is equal to
0.

By subtracting ${\mu_k}s_1^*(t)$, (\ref{Eq:y1}) can be further
written as
\begin{align}
    y'_{1,k}(t)&\triangleq{y_{1,k}(t)}-\mu_k{s_1(t)}
    \nonumber \\
    &={\nu_k}s_2^*(t-1)c_2^*(t)+w_{1,k}(t)
    \nonumber \\
            &=\left(y'_{1,k}(t-1)-w_{1,k}(t-1)\right)c_2^*(t)+w_{1,k}(t).
 \label{Eq:y1d1}
\end{align}

Finally, the following linear detector can be applied to recover
$c_2(t)$
\begin{align}
    \widetilde{c}_2&(t)=\text{arg}\underset{c_2(t)\in{\mathcal{A}}}{\max}\text{Re}\left\{y'^*_{1,k}(t-1){y'_{1,k}}(t)c_2(t)\right\}.
 \label{Eq:mldafk}
\end{align}

%
\subsection{Relay Selection with Differential Modulation in Two-Way Relay Channels}

In the proposed RS-ANC-DM scheme, only one best relay is selected
out of $N$ relays to forward the received ANC signals in the second
phase transmission. We assume that at the beginning of each
transmision, some pilot symbols are transmitted by two source nodes
to assist in the relay selection. One source node (either source
$S_1$ or $S_2$) will determine the one best relay according to a
certain criterion and broadcast the index of the selected relay to
all relays. Then only the selected relay is active in the second
phase of transmission and the rest of relays will keep idle.

\subsubsection{Optimal Single Relay Selection}

For the optimal single relay selection, a single relay node, which
minimizes the sum SERs of two source nodes, i.e.
$\text{SER}_{1,k}+\text{SER}_{2,k}$, will be selected, where
$\text{SER}_{1,k}$ and $\text{SER}_{2,k}$ represent the SERs at
source nodes $S_1$ and $S_2$, respectively. The main challenges in
relay selection for differential modulation as mentioned before is
that the relay node is determined by only one source (either $S_1$
or $S_2$) without the knowledge of any CSI. For simplicity and
without loss of generality, in the next we use $S_1$ to select the
optimal relay node. Obviously, the main difficulty here is to
estimate $\text{SER}_{2,k}$.

For $M$-PSK constellations, the conditional SER assoicated with the
$k$-th relay at the source $S_1$ is given by
~\cite{Simon-Digital-Comms}
\begin{equation}
    \text{SER}_{1,k}(h_{1,k},h_{2,k})
    =
    \frac{1}{\pi}\int_0^{\frac{(M-1)\pi}{M}}\exp\left(-\frac{g_\texttt{psk}\gamma_{d_1,k}}{\sin^2\theta}\right)d\theta,
 \label{Eq:conSER1}
\end{equation}
where $\gamma_{d1,k}$ is the effective SNR at the source $S_1$ and
$g_\texttt{psk}\triangleq\sin^2\frac{\pi}{M}$. As CSI is unknown to
the receiver, the effective SNR $\gamma_{d1,k}$ has to be estimated
without knowledge of CSI.

By ignoring the second order term, the corresponding
SNR of the proposed differential detection scheme in (\ref{Eq:y1d1})
can be written as
\begin{align}
    \gamma_{d_1,k}
    &\approx\frac{|{\nu_k}|^2}{\texttt{Var}\{2w_{1,k}(t)\}}
    \nonumber \\
    &\approx\frac{{\beta_k^2}|h_{1,k}|^2|h_{2,k}|^2}{2{\beta_k^2}N_0|h_{1,k}|^2+2N_0}
    \nonumber \\
    &\approx\frac{\psi_{r}\psi_{s}|h_{1,k}|^2|h_{2,k}|^2}
    {\psi_{r}|h_{1,k}|^2+\psi_{s}|h_{2,k}|^2},
 \label{Eq:SNRy1}
\end{align}
where
$\texttt{Var}\{w_{1,k}(t)\}\approx{\beta^2}N_0|h_{1,k}|^2+N_0$,
$\psi_{s}\triangleq{\frac{1}{4N_0}}$, and
$\psi_{r}\triangleq{\frac{1}{2N_0}}$.

Recalling $\mu_k\triangleq{\beta_k}|h_{1,k}|^2$,
$\nu_k\triangleq{\beta_k}h_{1,k}h_{2,k}^*$, and their corresponding
estimates in (\ref{Eq:nu}) and (\ref{Eq:muf}), $\gamma_{d1,k}$ in
(\ref{Eq:SNRy1}) can be further calculated as
\begin{align}
    \gamma_{d_1,k}\approx\frac{|\mu_k|^4|{\nu_k}|^2}{2(2|\mu_k|^2+|\nu_k|^2)(|\mu_k|^2+|\nu_k|^2)N_0},
 \label{Eq:SNRy1est}
\end{align}

Similar to (\ref{Eq:SNRy1est}), the SNR of the proposed differential
detection scheme in the source $S_2$ can be written as
\begin{align}
    \gamma_{d_2,k}
    &\approx\frac{\psi_{r}\psi_{s}|h_{1,k}|^2|h_{2,k}|^2}
    {\psi_{r}|h_{2,k}|^2+\psi_{s}|h_{1,k}|^2}
    \nonumber\\
    &\approx\frac{|\mu_k|^4|{\nu_k}|^2}{2(2|\nu_k|^2+|\mu_k|^2)(|\mu_k|^2+|\nu_k|^2)N_0}.
 \label{Eq:SNRy2}
\end{align}
And its SER can be then calculated as
\begin{equation}
    \text{SER}_{2,k}(h_{1,k},h_{2,k})
    =
    \frac{1}{\pi}\int_0^{\frac{(M-1)\pi}{M}}\exp\left(-\frac{g_\texttt{psk}\gamma_{d_2,k}}{\sin^2\theta}\right)d\theta.
 \label{Eq:conSER1}
\end{equation}

Among all relays, the destination will select one relay, denoted by
$\mathcal{R}$, which has the minimum destination SER:
\begin{align}
    \mathcal{R}=\underset{k}{\min}\left\{\text{SER}_{1,k}(h_{1,k},h_{2,k})+\text{SER}_{2,k}(h_{1,k},h_{2,k})\right\},
    k\in{1,\ldots,N}
 \label{Eq:relayselc}
\end{align}

\subsubsection{Sub-Optimal Single Relay Selection}
The optimal single relay selection scheme described in the above
section is very difficult to analyze. In this section we propose a
sub-optimal single relay selection scheme. It is well-known that the
sum SERs of two source nodes ($\text{SER}_{1,k}+\text{SER}_{2,k}$)
is typically dominated by the SER of the worst user. As a result,
for low complexity, the relay node, which minimizes the maximum SER
of two users, can be selected to achieve the near-optimal SER
performance. We refer to such a selection criterion as the Min-Max
selection criterion. Let $\mathcal{R}$ denote the selected relay.
Then the Min-Max selection can be formulated as follows,
\begin{align}
    \mathcal{R}=\underset{k}{\min}\max\left\{\text{SER}_{1,k}(h_{1,k},h_{2,k}),\text{SER}_{2,k}(h_{1,k},h_{2,k})\right\},k\in{1,\ldots,N},
 \label{Eq:relayselc}
\end{align}
which can be further formulated by using the effective SNRs
\begin{align}
    \mathcal{R}=\underset{k}{\max}\min\{\gamma_{d_1,k},\gamma_{d_2,k}\},k\in{1,\ldots,N},
 \label{Eq:relayselcsnr}
\end{align}
where the calculation of $\gamma_{d_1,k}$ and $\gamma_{d_2,k}$ can
be obtained from (\ref{Eq:SNRy1est}) and (\ref{Eq:SNRy2}),
respectively.

\section{Performance Analysis of RS-ANC-DM scheme Based on Min-Max Selection Criterion}
In this section, we derive the analytical average SER of the
proposed differential bi-directional relay selection schemes. As
mentioned before, the optimal relay selection scheme is very
difficult to analyze. As it will be shown later, the Min-Max
selection scheme proposed in section II has almost the same
performance as the optimal selection scheme. Therefore, in this
section, we will analyze the RS-ANC-DM scheme based on the Min-Max
selection criterion.
\subsection{Sub-Optimal Relay Selection}
For the Min-Max selection criterion, the effective SNR of the
selected relay $\mathcal{R}$ can be expressed as follows,
\begin{align}
    \gamma_{\mathcal{R}}={\max}\min\{\gamma_{d_1,k},\gamma_{d_2,k}\},k\in{1,\ldots,N}.
 \label{Eq:relayselcsnr}
\end{align}

Now let us first calculate the PDF of $\gamma_{\mathcal{R}}$. As
$\gamma_{d_1,k}$ and $\gamma_{d_2,k}$ are identically distributed,
they have the same PDF and CDF, denoted by $f_{\gamma_{k}}(x)$ and
$F_{\gamma_{k}}(x)$, respectively.

Define
$\gamma_k^{\min}{\triangleq}\min\{\gamma_{d_1,k},\gamma_{d_2,k}\}$.
Let $f_{\gamma_k^{\min}}(x)$ and $F_{\gamma_k^{\min}}(x)$ represent
its PDF and CDF, respectively. For simplicity, assuming that $\gamma_{d_1,k}$ and $\gamma_{d_2,k}$ are independent, then the PDF of $\mathcal{R}$ can be
calculated by using order statistics as \cite{David1970}
\begin{align}
    f_{\gamma_{\mathcal{R}}}(x)=Nf_{\gamma_k^{\min}}(x)F_{\gamma_k^{\min}}^{N-1}(x)
            =2Nf_{\gamma_{k}}(x)(1-F_{\gamma_{k}}(x))[1-(1-F_{\gamma_{k}}(x))^2]^{N-1},
 \label{Eq:fpdff}
\end{align}
where
$f_{\gamma_k^{\min}}(x)=2f_{\gamma_{k}}(x)(1-F_{\gamma_{k}}(x))$,
$F_{\gamma_k^{\min}}(x)=1-(1-F_{\gamma_{k}}(x))^2$, and
$f_{\gamma_{k}}(x)$ can be found in~\cite{Hasna}
\begin{align}
    f_{\gamma_{k}}(x)=
    \frac{2x\exp\left(-x(\psi_{r}^{-1}+\psi_{s}^{-1})\right)}{\psi_{r}\psi_{s}}
    \left[\frac{\psi_{r}+\psi_{s}}{\sqrt{\psi_{r}\psi_{s}}}
            \right.
    {\times}K_1\left(\frac{2x}{\sqrt{\psi_{r}\psi_{s}}}\right)
    \vspace{-3em}\left.+2K_0\left(\frac{2x}{\sqrt{\psi_{r}\psi_{s}}}\right)\right]U(x),
 \label{Eq:PDFaafk}
\end{align}
where $K_0(\cdot)$ and $K_1(\cdot)$ denote the zeroth-order and
first-order modified Bessel functions of the second kind,
respectively, and $U(\cdot)$ is the unit step function. At high SNR,
when $z$ approaches zeros, the $K_1(z)$ function converges to
$1/z$~\cite{Abramowitz}, and the value of the $K_0(\cdot)$ function
is comparatively small, which could be ignored for asymptotic
analysis. Hence, at high SNR, $f_{\gamma_{k}}(x)$ in
(\ref{Eq:PDFaafk}) can be reduced as
\begin{align}
    f_{\gamma_{k}}(x)\approx
    \frac{\psi}{2}\exp\left(-\frac{\psi}{2}x\right),
 \label{Eq:PDFPx}
\end{align}
where $\psi\triangleq{2}({\psi_{r}^{-1}+\psi_{s}^{-1}})$. Its
corresponding CDF can be written as
\begin{align}
    F_{\gamma_{k}}(x)\approx1-\exp\left(-\frac{\psi}{2}x\right).
 \label{Eq:CDFPx}
\end{align}

The PDF of $\gamma_{\mathcal{R}}$ can then be approximately calculated as
\begin{align}
    f_{\gamma_{\mathcal{R}}}(x)\approx{N}\psi\exp\left(-\psi{x}\right)
    \left[1-\exp(-\psi{x})\right]^{N-1}.
 \label{Eq:fpdf}
\end{align}

The CDF of $\gamma_{\mathcal{R}}$ can be approximately written as
\begin{align}
    F_{\gamma_{\mathcal{R}}}(x)\approx\left[1-\exp(-\psi{x})\right]^{N}
    \approx\left(\psi{x}\right)^{N},
 \label{Eq:fcdf}
\end{align}
where $\underset{x\rightarrow{\infty}}\lim1-\exp(-x)=x$.

The SER conditioned on the instantaneous received SNR is
approximately \cite{Proakis-Digital-Comms}
\begin{align}
    \text{SER}(\gamma_{\mathcal{R}}|h_{1},h_{2})\approx{Q\left(\sqrt{c\gamma_{\mathcal{R}}}\right)},
 \label{Eq:inSER}
\end{align}
where $Q(\cdot)$ is the Gaussian-$Q$ function, $Q(x)=\frac{1}{\sqrt{2\pi}}\int_x^\infty\exp(-t^2/2)\text{d}t$, and $c$ is a constant determined
by the modulation format, e.g. $c=2$ for BPSK constellation.

The average SER can be then derived by averaging over the Rayleigh fading channels by
\begin{align}
    \text{SER}(\gamma_{\mathcal{R}})=\mathbb{E}\left[\text{SER}(\gamma_{\mathcal{R}}|h_{1},h_{2})\right]
    =\mathbb{E}\left[{Q\left(\sqrt{c\gamma_{\mathcal{R}}}\right)}\right].
 \label{Eq:aveSERtemp}
\end{align}

By introducing a new random variable (RV) with standard Normal distribution $X\sim{\mathcal{N}(0, 1)}$, the average SER can be rewritten as \cite{Zhao09}
\begin{align}
    \text{SER}(\gamma_{\mathcal{R}})&=P\left\{X>\sqrt{c\gamma_{\mathcal{R}}}\right\}
    \nonumber \\
    &=P\left\{\gamma_{\mathcal{R}}<\frac{X^2}{c}\right\}
    \nonumber \\
    &=\mathbb{E}\left[F_{\gamma_{\mathcal{R}}}\left(\frac{X^2}{c}\right)\right]
    \nonumber \\
    &=\frac{1}{\sqrt{2\pi}}\left(\frac{\psi}{c}\right)^{N}\int_0^\infty{x^{2N}}\exp\left(-\frac{x^2}{2}\right)\text{d}x.
 \label{Eq:aveSERre}
\end{align}
Based on the fact that
$\int_0^\infty{t^{2n}}\exp(-kt^2)\text{d}t=\frac{(2n-1)!!}{2(2k)^n}\sqrt{\frac{\pi}{k}}$
\cite{Gradshteyn94}, we can finally obtain
\begin{align}
    \text{SER}(\gamma_{\mathcal{R}})=\frac{(2N-1)!!}{2}\left(\frac{\psi}{c}\right)^{N},
 \label{Eq:aveSER}
\end{align}
where $(2n-1)!!=\prod_{k=1}^{n}{2k-1}=\frac{(2n-1)!}{n!2^n}$.

It clearly indicates in (\ref{Eq:aveSER}) that a diversity order of
$N$ can be achieved for the proposed RS-ANC-DM scheme in a
bi-directional relay network with two sources and $N$ relays.

\section{Simulation Results}
In this section, we provide simulation results for the proposed
RS-ANC-DM scheme. We also include the corresponding coherent
detection results for comparison. All simulations are performed with
a BPSK modulation over the Rayleigh fading channels, and the frame
length is $L=100$. In order to calculate $\nu_k$ in (\ref{Eq:nu}),
we use the normal constellation and the constellation rotation
approaches introduced in Appendix~A. For simplicity, we assume that
$S_1$, $S_2$, and $R_k$ ($k={1,\ldots,N}$) have the same noise
variance.
\subsection{Simulated Results}
\begin{figure}[]
\centering
\includegraphics[height=4.2in,width=5.5in]{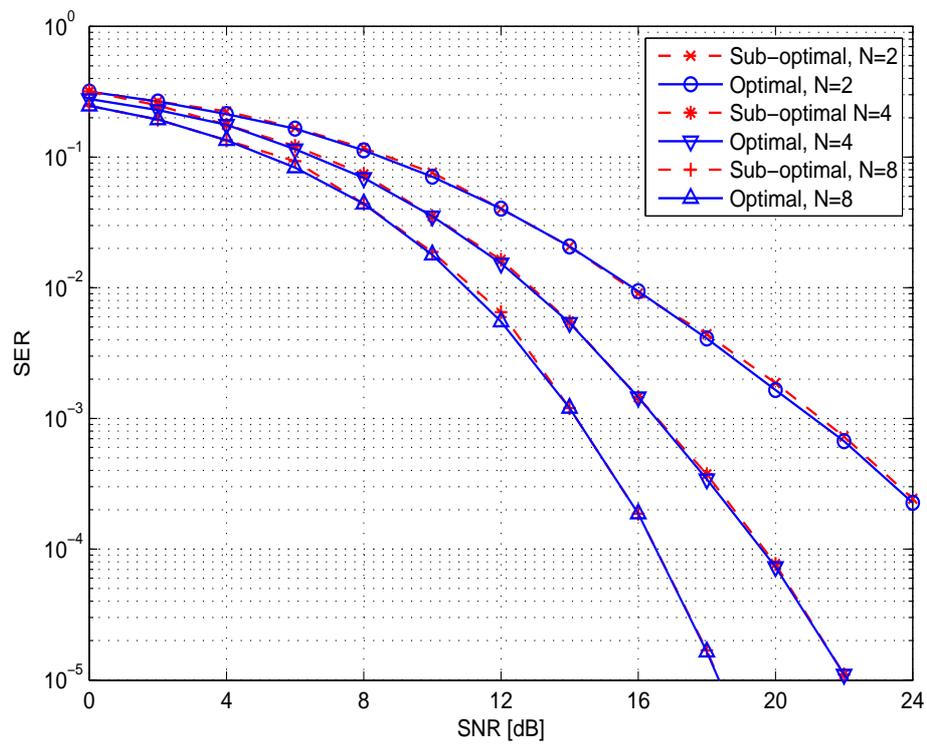}
\caption{Simulated SER performance by optimal and sub-optimal
detections.} \label{Fig:DBRS_Opt_SubOpt}
\end{figure}

\begin{figure}[]
\centering
\includegraphics[height=4.2in,width=5.5in]{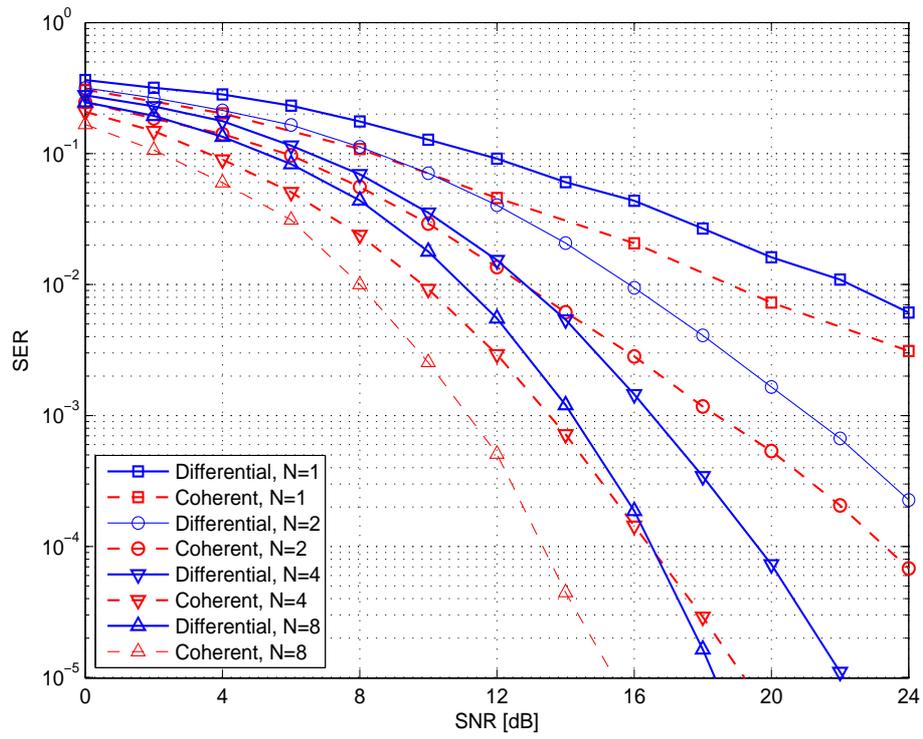}
\caption{Simulated SER performance by differential and coherent
detections, where $N=1,2,4,8$.} \label{Fig:Coh_diff_BRS}
\end{figure}

\begin{figure}[]
\centering
\includegraphics[height=4.2in,width=5.5in]{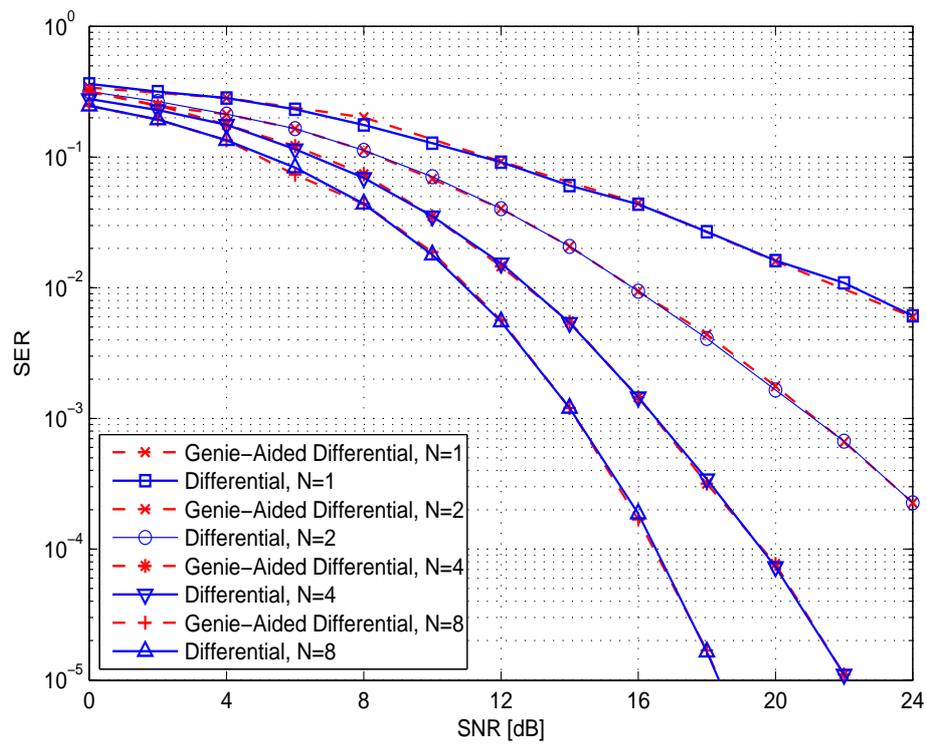}
\caption{Simulated SER performance by differential and Genie-aided
detections, where $N=1,2,4,8$.} \label{Fig:DBRS_SubOpt_Genie}
\end{figure}

From Fig.~\ref{Fig:DBRS_Opt_SubOpt} to
Fig.~\ref{Fig:DBRS_SubOpt_Genie}, we present the simulated SER
performance for the proposed RS-ANC-DM schemes. The performance of
the corresponding coherent detection are plotted as well for better
comparison. In Fig.~\ref{Fig:DBRS_Opt_SubOpt}, we compare the
optimal relay selection method and the sub-optimal Min-Max relay
selection method. It can be observed from the figure that the
proposed Min-Max selection approach has almost the same SER as the
optimal one. In particular, when the number of relay nodes
increases, we almost cannot observe any difference between these two
methods, which indicates that the Min-Max relay selection achieves
near optimal single relay selection performance.

In Fig.~\ref{Fig:Coh_diff_BRS}, we compare the RS-ANC-DM scheme with
its coherent detection counterpart. It can be noted that the
differential scheme suffers about 3 dB performance loss compared to
the corresponding coherent scheme. We can also see from
Fig.~\ref{Fig:Coh_diff_BRS} that the SER performance is
significantly improved when the number of the relay increases. In
Fig.~\ref{Fig:DBRS_SubOpt_Genie}, we include the Genie-aided result
by assuming that $\mu$ is perfectly known by the source such that
traditional differential decoding can be performed. It shows from
the results that there is almost no performance loss using the
estimation method in (\ref{Eq:muf}) which clearly justifies the
robustness of the proposed differential decoder.

\begin{figure}[]
\centering
\includegraphics[height=4.2in,width=5.5in]{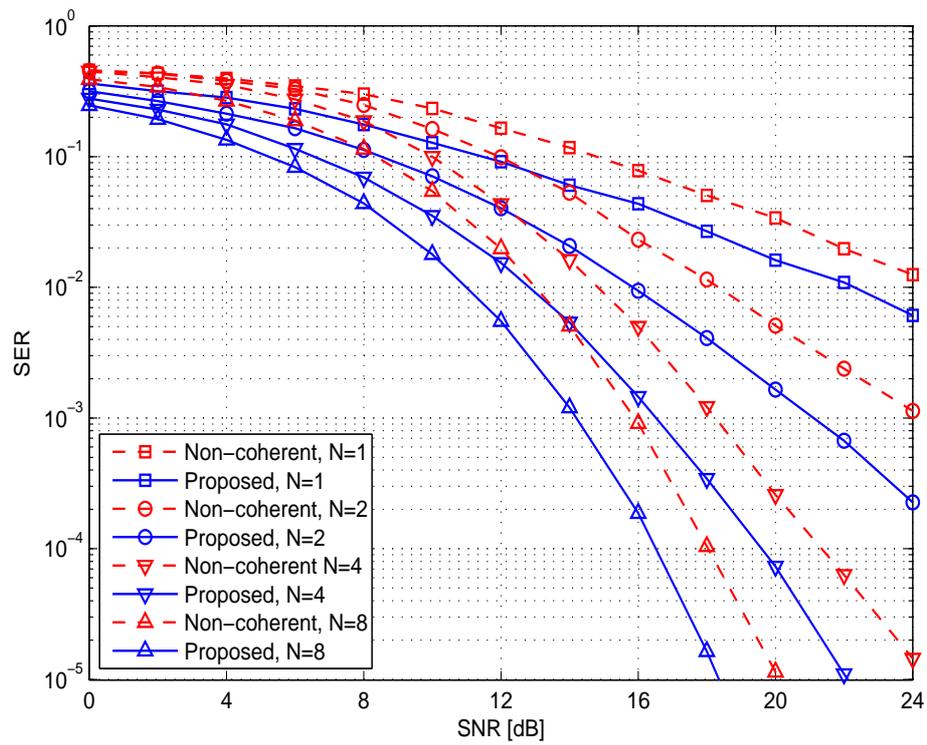}
\caption{Simulated SER performance comparisons by our proposed
differential approach and the non-coherent scheme in~\cite{Tao2008},
where $N=1,2,4,8$.} \label{Fig:Diff_bidirectional_relay}
\end{figure}

Fig.~\ref{Fig:Diff_bidirectional_relay} compares the simulated SER
performance for our proposed RS-ANC-DM and the non-coherent
schemes~\cite{Tao2008} in bi-directional relaying without using
constellation rotation, where $N=1,2,4,8$. It can be observed that
our proposed scheme has much better performance than the detector in
\cite{Tao2008}. The main reason is that the non-coherent detection
approach employed in \cite{Tao2008} statistically averages off the
impact of channel fading coefficients by ignoring the instantaneous
channel state information and thus causes much performance loss.
Comparatively, our proposed differential detection is a symbol by
symbol based detection and is thus be able to adapt to the variation
of the channel.
\subsection{Analytical Results}
\begin{figure}[]
\centering
\includegraphics[height=4.2in,width=5.5in]{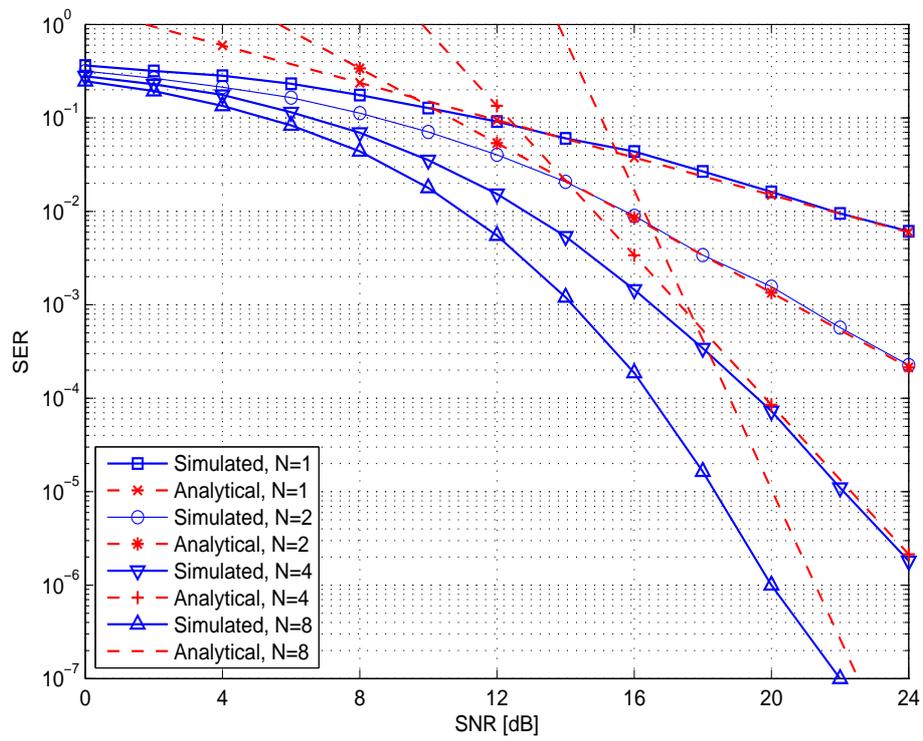}
\caption{Analytical and Simulated SER performance by the proposed
differential scheme, where $N=1,2,4,8$.} \label{Fig:DBRS_Ana}
\end{figure}

In Fig.~\ref{Fig:DBRS_Ana}, we compare the analytical and simulated
SER performance of the proposed differential modulation scheme.I n
order to obtain fine estimation in (\ref{Eq:nu}), the signal
constellation used by $S_1$ is rotated by $\pi/2$ relative to that
by $S_2$. From the figure, it can be observed that at high SNR, the
analytical SER derived by (\ref{Eq:aveSER}) is converged to the
simulated result using optimal relay selection. This verifies the
derived analytical expressions.
\subsection{Constellation Rotation}
\begin{figure}[]
\centering
\includegraphics[height=4.2in,width=5.5in]{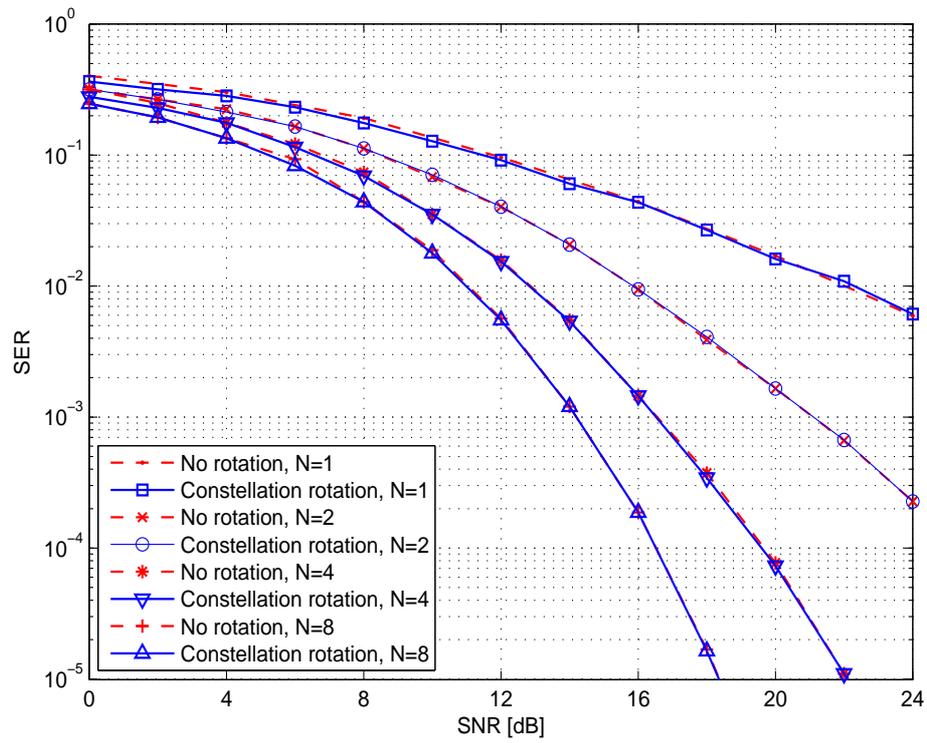}
\caption{Simulated SER performance by the proposed differential
scheme with and without using constellation rotation, where
$N=1,2,4,8$.} \label{Fig:DBRS_Constellation_Rotation}
\end{figure}

In Fig~\ref{Fig:DBRS_Constellation_Rotation}, we examine the SER
results of the proposed differential modulation scheme in comparison
with the one without using constellation rotation, as shown in
Appendix~A, where the signal constellation used by $S_1$ is rotated
by $\pi/2$ relative to that by $S_2$. It can be observed that the
new result has very similar with the curve without rotating
constellations. This indicates that using constellation rotation may
not obtain any gains given large frame length.
\section{Conclusions}
In this paper, we proposed a joint relay selection and analog
network coding using differential modulation over two-way
relay channels when neither sources nor the relay has access to the
channel state information. A simple Min-Max relay selection method
is proposed and it has been shown that it achieves almost the same
performance as the optimal single relay selection scheme. An
asymptotic SER expression is derived. It is shown that the proposed
RS-ANC-DM scheme can achieve the full diversity order of $N$ for the system
with $N$ relays. Results are verified through simulations.
%
\appendices
\section{The Calculation of $\mathbb{E}[|c_1(t)-c_2(t)|^2]$ in (\ref{Eq:nu})}
From (\ref{Eq:nu}), it shows that the average power of
$c_1(t)-c_2(t)$ needs to be calculated. When $M$-PSK constellations
are applied, the number of symbols produced in the new
constellations by $c_1(t)-c_2(t)$ is finite. Hence, it is easy to
derive the average power of the new constellation sets. We refer to
this as the normal constellation approach.

Note that the value of $c_1(t)-c_2(t)$ can be equal to zero, which
may affect the estimation accuracy in (\ref{Eq:nu}). In order to
overcome this problem, we may properly choose a rotation angle for
the symbol modulated in source $S_2$ by $c_2(t)e^{-j\theta}$,
ensuring that $c_1(t)-c_2(t)$ in (\ref{Eq:nu}) is nonzero. For a
$M$-PSK constellation, the effective rotation angle is in the
interval $[-\pi/M,\pi/M]$ from the symmetry of symbols. For a
regular and symmetrical constellation, the rotation angle may be
simply set as $\theta=\pi/M$. Similar approach may be used to
generate the rotation angle for other types of constellations.

\pagebreak
\clearpage
\clearpage
\clearpage
\clearpage
\clearpage
\clearpage
\clearpage
\clearpage

\end{document}